\documentclass[useAMS,usenatbib,usegraphicx]{mn2e}
\usepackage{graphicx}
\DeclareGraphicsExtensions{.pdf,.png,.jpg,.eps}
\usepackage{tabularx}
\usepackage{amsmath}
\usepackage{txfonts}
\usepackage{wasysym}
\usepackage{ragged2e}
\usepackage{natbib}
\usepackage{aas}
\usepackage[english]{babel}
\newcommand{\hst}{\textsl{HST}}
\usepackage{float}

\title[The 
M\,4 Core Project with \hst -- III. Search for variable stars in the primary field]{The 
M\,4 Core Project with \hst -- III. Search for variable stars in the primary field\thanks{Based 
on observations collected with the NASA/ESA Hubble Space Telescope, obtained at the Space 
Telescope Science Institute, which is operated by AURA, Inc., under NASA contract 
NAS 5-26555, under large program GO-12911.}}

\author[V.\ Nascimbeni et al.]{V.\ Nascimbeni$^{1}$\thanks{Corresponding authors: e-mail:
valerio.nascimbeni@unipd.it (VN); luigi.bedin@oapd.inaf.it (LRB)},
L.\ R.\ Bedin$^1$,
D.\ C.\ Heggie$^2$,
M.\ van den Berg$^{3,4}$,  
M.\ Giersz$^5$, 
G.\ Piotto$^{1,6}$, \newauthor  
K.\ Brogaard$^{7,8}$,  
A.\ Bellini$^9$, 
A.\ P.\ Milone$^{10}$, 
R.\ M. Rich$^{11}$,   
D.\ Pooley$^{12,13}$, 
J.\ Anderson$^9$, \newauthor 
L.\ Ubeda$^9$, 
S.\ Ortolani$^{1,6}$,  
L.\ Malavolta$^{1,6}$, 
A.\ Cunial$^{1,6}$, and 
A.\ Pietrinferni$^{14}$ \\
$^{1}$INAF - Osservatorio Astronomico di Padova, Vicolo dell'Osservatorio 5, Padova, IT-35122 \\
$^{2}$School of Mathematics and Maxwell Institute for Mathematical Sciences, University of Edinburgh, Kings Buildings, Edinburgh, UK-EH9-3JZ \\
$^{3}$Anton Pannekoek Institute for Astronomy, University of Amsterdam, Science Park 904, 1098 XH Amsterdam, The Netherlands \\
$^{4}$Harvard-Smithsonian Center for Astrophysics, 60 Garden Street, Cambridge, 02138 MA, USA \\
$^{5}$Nicolaus Copernicus Astronomical Center, Polish Academy of Sciences, ul.\ Bartycka 18, 00-716, Warsaw, Poland \\
$^{6}$Dipartimento di Fisica e Astronomia ``Galileo Galilei'', Universit\`a di Padova, Vicolo dell'Osservatorio 3, Padova IT-35122 \\
$^{7}$Stellar Astrophysics Centre, Department of Physics and Astronomy, Aarhus University, Ny Munkegade, 8000 Aarhus C, Denmark\\ 
$^{8}$Department of Physics and Astronomy, University of Victoria, PO Box 3055, Victoria, B.C., V8W 3P6, Canada \\
$^{9}$Space Telescope Science Institute, 3800 San Martin Drive, Baltimore, MD 21218, USA\\
$^{10}$Research School of Astronomy and Astrophysics, The Australian National University, Cotter Road, Weston, ACT, 2611, Australia \\
$^{11}$Department of Physics and Astronomy, University of California, Los Angeles, CA 90095, USA \\
$^{12}$Department of Physics, Sam Houston State University, Huntsville, TX 77341, USA \\
$^{13}$Eureka Scientific, Inc., 2452 Delmer Street, Suite 100, Oakland, CA 94602, USA \\
$^{14}$INAF - Osservatorio Astronomico di Teramo, Via M. Maggini, 64100 Teramo, Italy.} 
\begin{document}

\date{Accepted May 7, 2014. Submitted Apr 5, 2014. [Compiled \today]}

\pagerange{\pageref{firstpage}--\pageref{lastpage}} \pubyear{2002}

\maketitle

\label{firstpage}

\begin{abstract}
We present the results of a photometric search for variable stars 
in the core of the Galactic globular cluster M\,4. 
The input data are a large and unprecedented set of deep Hubble Space Telescope WFC3 
images (large program GO-12911; 120 orbits allocated), primarily aimed at probing binaries with massive 
companions by detecting their astrometric wobbles. Though these data were not optimised to carry out
a time-resolved photometric survey, their exquisite precision, spatial resolution and dynamic
range enabled us to firmly detect 38 variable stars, of which 20 were previously unpublished. They
include 19 cluster-member eclipsing binaries (confirming the large binary fraction of M\,4), RR Lyrae, 
and objects with known X-ray counterparts. We improved and revised the parameters 
of some among published variables. 
\end{abstract}

\begin{keywords}
globular clusters: individual: NGC 6121 --
stars: variables: general --
binaries: general --
techniques: photometric.
\end{keywords}

%%%%%%%%%%%%%%%%%%%%%%%%%%%%%%%%%%%%%%%%%%%%%%%%%%
\section{Introduction}
%%%%%%%%%%%%%%%%%%%%%%%%%%%%%%%%%%%%%%%%%%%%%%%%%%

Messier\ 4 (M\,4), also known as NGC\,6121 is the closest Galactic globular cluster (GC)
at 1.86 kpc, having the
second smallest apparent distance modulus after NGC\,6397: $(m-M)_V = 12.68$ \citep{bedin2009}. 
It is known to show no evidence for any central brightness cusp, despite being
significantly older than its dynamical relaxation time \citep{trager1995}. Further, the photometric
binary fraction in the core of M\,4 is among the highest measured for a GC, reaching 15\% in
the core region (\citealt{milone2012}; compare with 2\% for NGC\,6397). The fine details of the
role played by dynamical interactions between binary stars in the delay of cluster
``core collapse'' are still debated, with different competing theories proposed 
to explain them (see \citealt{heggie2003} for a review). As M\,4 appears 
to be a perfect case to test those theories, we proposed a Hubble Space Telescope
(\textit{HST}) large program entitled ``A search for binaries with massive companions in the core of
the closest globular cluster M\,4'' (GO-12911, PI: Bedin), which has been awarded 
120 orbits and has already been successfully completed. The main aim of our project 
is to probe that fraction of binary population which is undetectable by means of 
usual photometric techniques, 
viz.~the fraction that is made up of binaries composed of
a main-sequence (MS) star and
a massive, faint evolved companion (e.g., black hole, white dwarf, neutron star).
The employed technique is an astrometric search for wobbles due to the motion of the 
bright component around the system barycentre. In support of this program, a set of 720
WFC3/UVIS (Wide Field Camera 3, Ultraviolet and VISual channel)
images have been gathered over a baseline of about one year. The UVIS 
$162''\times 162''$ field of view (FOV) covers the whole core of
M\,4 (whose radius is $r_c\simeq 70''$; \citealt{harris1996}) at 
every roll angle.
Of course,
such a massive data set can be exploited for a large number of tasks other than the
primary one. We will refer the reader to Paper I \citep{bedin2013} for a detailed
description of the program and for a discussion about other possible collateral 
science.

M\,4 is a cluster which has some desirable properties for
photometric searches of variables among population II stars, 
% RMR & VDB
viz.~both its proximity and relatively low density core.
For this reason it has been intensively targeted since the 
beginning of the photographic and photoelectric era 
\citep{leavitt1904,sawyer1931,greenstein1939, desitter1947}
up to more recent works 
which employed CCD photometry, either ground-based \citep{kaluzny1997,kaluzny2013b}
or space-based \citep{ferdman2004}. Besides a large number of known RR Lyrae
(53, according to the online 
database\footnote{http://www.astro.utoronto.ca/\textasciitilde{}cclement/cat/listngc.html 
(\citealt{clement2001},
last update 2009).} compiled by C.~Clement), eclipsing binaries and other 
ordinary variables, M\,4 also hosts other less common objects of interest, 
including an exotic planetary system made of a pulsar, a white dwarf, 
and a 2.5-$M_\mathrm{jup}$ planet \citep{sigurdsson2003}
and many X-ray sources \citep{bassa2004} whose optical 
counterparts show both periodic or irregular photometric variability 
\citep{kaluzny2012}.

In this study we exploit the GO-12911 data set to
extract 9\,410 light curves of every well-measured, point-like source 
in the M\,4 core, spanning from the horizontal branch (HB) down to 
the lower main sequence (MS). 
We aim at discovering new 
variable stars and at refining the parameters of some others that were previously published;
this includes the firm identification of a few objects for which 
the physical nature and/or the cluster membership was
classified as ``uncertain'' in the past. We describe in Section 2 
the criteria adopted to select the input list, the procedures to correct the 
light curves by means of differential local photometry,
and the specific algorithms employed to perform the search for periodic 
and irregular variability and to sift the most significant candidates. 
Then, in Section 3, we present our list of 38 high-confidence variables,
along with a discussion of some notable individual cases.  
The overall statistics of our set, and in particular about its
completeness limits and biases, is eventually discussed in the Section\,4.

\begin{figure*}
\includegraphics[width=\columnwidth]{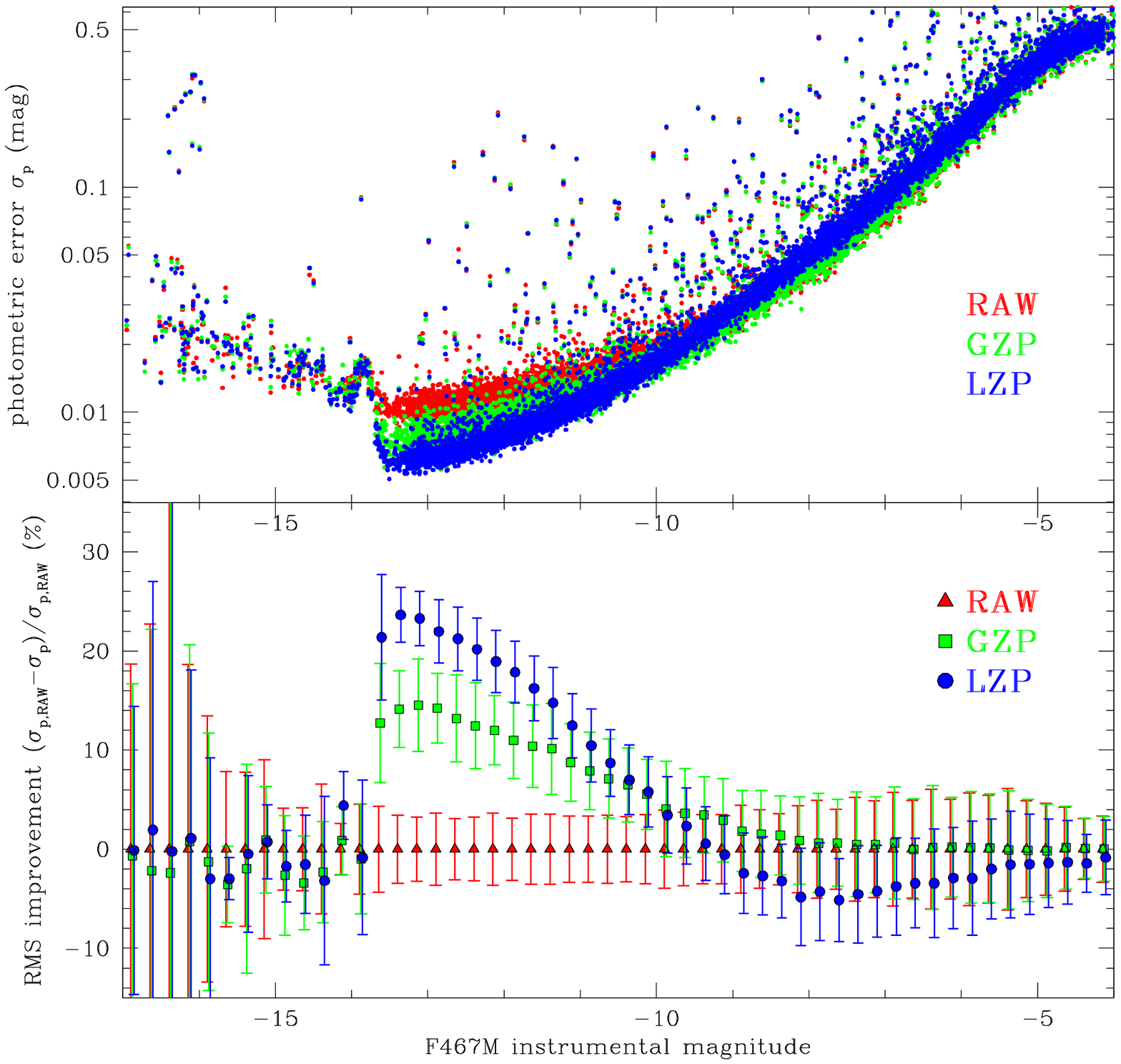}
\includegraphics[width=\columnwidth]{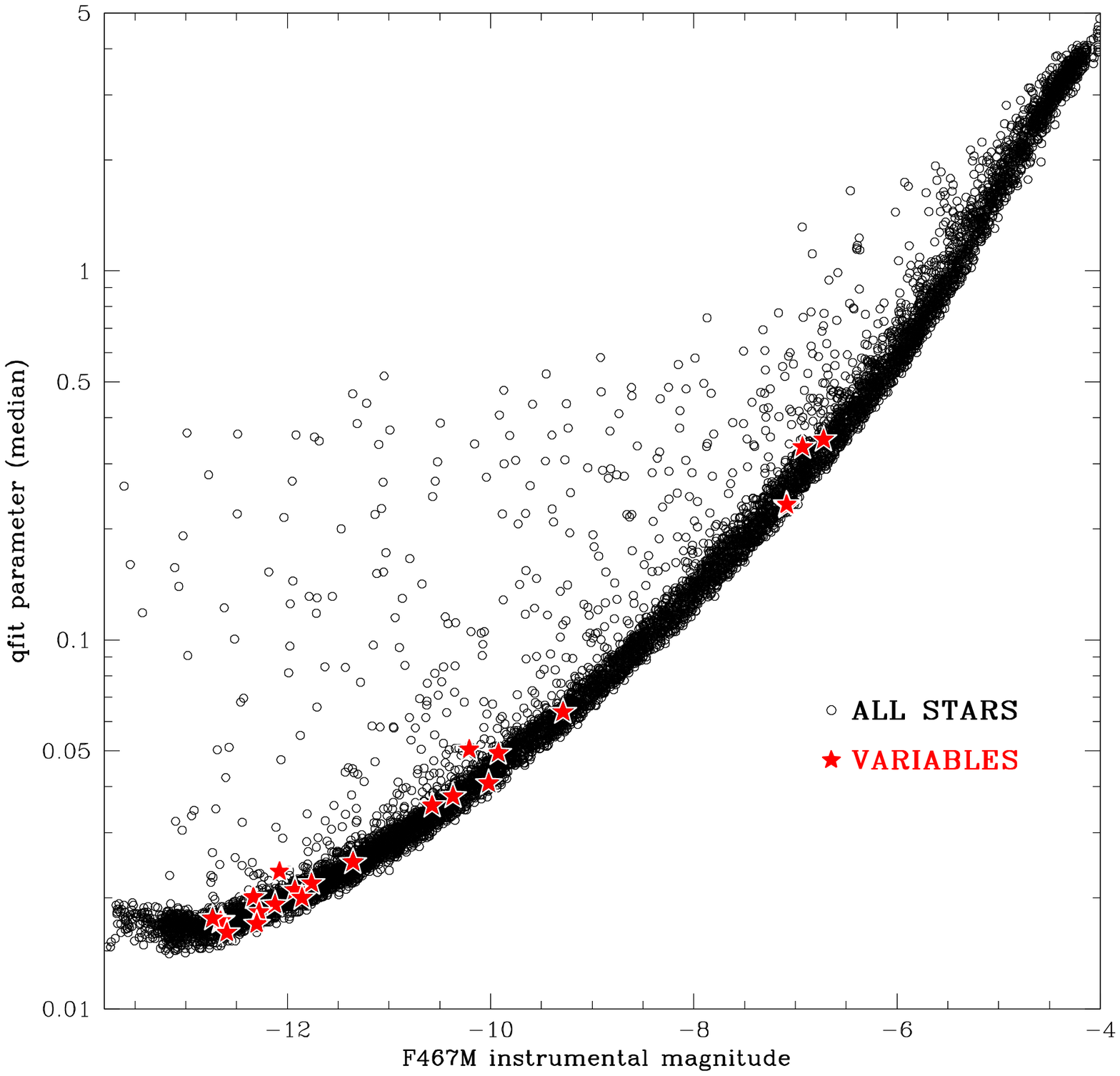}
\caption{\emph{Left, upper panel: } Photometric error $\sigma_p$ as a function of instrumental magnitude F467M,
measured on all stars with three different correction algorithms: none (``RAW'', red points), global zero-point
(``GZP'', green) and local zero-point (``LZP'', blue). Saturation occurs at $\mathrm{F467M} \simeq -13.75$.
\emph{Left, lower panel:} ``improvement'' as a function of F467M, i.e., the percent reduction of RMS 
compared to raw light curves, averaged on 0.25-mag bins. Colours are coded as above. \emph{Right:} median
value of the fit quality parameter \texttt{qfit} as a function of F467M, all light curves (black circles). 
Variable stars found by our study (Table \ref{variables}) are marked with red stars.}
\label{rmsq}
\end{figure*}

%%%%%%%%%%%%%%%%%%%%%%%%%%%%%%%%%%%%%%%%%%%%%%%%%%
\section{Data analysis} 
%%%%%%%%%%%%%%%%%%%%%%%%%%%%%%%%%%%%%%%%%%%%%%%%%%

The full GO-12911 data set was gathered during 120 HST orbits,
arranged in 12 epochs made of 10 \textit{HST\/} visits each, where each visit is 
one orbit. 
Each orbit is filled with five $392$-$396$ s exposures
in the blue filter F467M (except for eleven isolated frames for which the F467M exposure time was set to
366 s), and one additional 20-s exposure through a red F775W filter at the beginning of
the orbit. 
The choice of such unorthodox filters was driven by the 
astrometric requirements of our project. The intermediate-band F467M yields a more 
monochromatic-like point spread function (PSF), less prone to colour-dependent systematic errors.
For the same reason, the Sloan $i'$-like F775W filter was preferred over the  more commonly used F814W
thanks to its better characterised astrometric solution. Its $\sim$$1000$ $\AA$ cut on the red 
tail does not imply a significant flux loss, because there the total transmission is very low.
The F467M filter also has the advantage of suppressing the contaminating light of PSF halos from red giants. 

The first visit occurred on 2012 Oct 9, followed by a 100 d gap and then by
eleven other visits regularly spaced at a $\sim$24-day cadence. In this work we 
will focus on a homogeneous subset of 589 ``deep'' ($392$-$396$ s) F467M images, 
in order to take advantage of a denser sampling and smaller flux contamination
from giants.
Data reduction was carried out by modelling an effective PSF
(ePSF; \citealt{anderson2000}) tailored on each frame. Details about the ePSF approaches 
can be found in Paper I and references therein.
Proper motions were derived by matching our data with ACS
astrometry by \citet{sarajedini2007} (GO-10775, PI: Sarajedini), over a baseline
of about six years.

It is worth noting that the codes employed are able
to extract acceptable photometry even on stars brighter than saturation
by collecting the charge bled
along the columns of the detector \citep{gilliland2004,anderson2008}.
This is possible thanks to the excellent 
capability of WFC3/UVIS to conserve the flux even after the pixel full-well 
is exceeded \citep{gilliland2010}. 
Each unsaturated star in each exposure was measured by adding up the
flux within its central $5\times 5$ pixels, then dividing by the fraction of
the star's light that should have fallen within the aperture (based on
the PSF model and the PSF-fitted position of the stars within its
central pixel).  The aperture for saturated stars started with this
$5\times 5$ aperture, but we also had to include all contiguous pixels that
were either saturated or neighboring saturated pixels.  The total flux
was then the flux of the star through the aperture divided by the
fraction of the PSF determined to lie within the aperture.  In this
way, we were able to determine the photometry of the saturated and
unsaturated stars in the same system.  Fig.~\ref{rmsq} shows that the
absolute precision of the saturated stars is not as good as that for
the unsaturated stars, since the LZP and GZP are constructed to
correspond to the $5\times 5$-pixel aperture, not the variable aperture for
saturated stars.  Nevertheless, the smoothness of our color-magnitude diagram
(CMD; Fig.~\ref{cmd}) across
the F467M saturation boundary at $-13.75$ indicates that there are no
systematic differences between photometry for the saturated and
unsaturated stars.  
The reliability of this approach is shown by the 
quality of the light curves for the 13 RR
Lyrae stars (see Sect.\,3.1 for details).

Our initial input list was constructed by requiring the detection for each given 
source in at least 100 out of  589 frames, in order to get light curves spanning a
phase coverage large enough to extract a meaningful period analysis from them. 
This constraint left us with 9\,410 sources, all brighter than instrumental 
magnitude $\mathrm{F467M}\simeq -4$, corresponding to about 40
detected photoelectrons. On the bright side, the most luminous stars
reach $\mathrm{F467M}\simeq -17.5$ (corresponding to $V\simeq12.5$: Fig.~\ref{rmsq}, upper left panel). This means that
the dynamic range of our data set spans more than 13 magnitudes, enabling us
to measure stars which are usually saturated and neglected in most surveys.

Most sources among our detections
are single, point-like sources belonging to M\,4.
A small fraction of the sample, however,  is made of galaxies, extended objects, 
unresolved stellar blends and instrumental artefacts. These can be
identified by looking at the \texttt{qfit} parameter, a
diagnostic value that is related to the goodness of the ePSF fit 
\citep{anderson2008}. 
As the mean $\langle\texttt{qfit}\rangle$ of a light curve is a monotonically increasing 
function of magnitude (Fig.~\ref{rmsq}, right panel), a reliable way
to identify badly-measured outliers is to compare it with its median value evaluated over
magnitude bins. We anticipate that, after the vetting procedure, all the
variable stars discussed here are high-confidence point-like sources.

The observing strategy behind the GO-12911 program was optimised 
for performing high precision astrometry. In order to model and correct all
the distortion terms of the astrometric solution, the images were 
gathered 
by setting a large-dither pattern of 50 points and changing
the telescope roll angle within each astrometric epoch.
Stars fall on completely different physical 
pixels on most frames. While this is a winning choice for the 
main goal of program, it poses 
some issues when trying to extract accurate time-resolved photometry. 
Even when the PSFs are carefully modelled, this approach amplifies the effect of
flat field residual errors, intrapixel and pixel-to-pixel 
inhomogeneities, and other position-dependent effects. As a 
consequence, subtle second-order systematics are introduced 
in our photometry, as is evident by examining the raw light 
curves which share common trends whose shape and amplitude 
depend on the sky region analysed. A similar behaviour was also 
observed on data from the ACS/WFC (Advanced Camera for Surveys, 
Wide Field Channel) in our previous 
work on NGC\,6397 \citep{nascimbeni2012}.
Our approach to minimise such systematics is based on 
correcting differential light curves by subtracting a local 
zero-point (LZP), evaluated individually for each target star
and each frame.

\subsection{Global differential photometry}\label{ssgzp}

Before performing a LZP correction, an intermediate 
and straightforward step is applying a
global zero-point correction (GZP).  
In what follows, we index the individual
frames with variable $i$, the 
9\,410
target stars with the variable $k$ 
and the subset of sources chosen as comparison stars 
with $j$. Individual data points from target/reference
light curves will be then identified by $m_{i,k}$ and $m_{i,j}$, respectively.
The notation $\langle x \rangle_y$ represents the averaging of $x$ 
over the index $y$. Unless otherwise noted, averaging is done by evaluating 
the median and setting as the associated scatter $\sigma_{\langle x\rangle}$
the $68.27^\mathrm{th}$ percentile of the absolute
residuals.

As a first step, we chose a common set of reference sources.
These are required to be bright, non-saturated ($-13.75<F467M<-10$), 
point-like and well-fitted (\texttt{qfit} within 2$\sigma$ from
the median \texttt{qfit} of all the stars having similar magnitude; this
ensures that extended sources and blends are discarded). We also required that they
are detected and measured on a minimum number of frames
$N_\mathrm{min}=500$ over 589, as a reasonable compromise between 
completeness and FOV coverage. 
This left us with 1\,485 reference stars.
For each of them we computed the median raw magnitude $\langle m_{i,j} \rangle_i$  
by iteratively clipping outliers at $2\sigma$; 
then each reference raw light curve $m_{i,j}$ was normalised by 
subtracting $\langle m_{i,j} \rangle_i$ from
it. A global ``trend'' $\tau_i$ was calculated for each frame $i$ by taking 
the 2-$\sigma$ clipped median of all available $m_{i,j}-\langle m_{i,j} \rangle_i$. 
The quantity $\tau_i$ is the GZP correction to be applied to each point 
$m_{i,k}$ belonging to target light curves: 
\begin{equation}\label{gzp}
 m'_{i,k} = m_{i,k} - \tau_i = m_{i,k} - \langle m_{i,j} - \langle m_{i,j} \rangle_i \rangle_j 
\textrm{ .}
\end{equation}

The pre-normalisation procedure enables us to estimate $\tau_i$ 
without biases, even if a
small subset of reference stars is lacking from a given frame. On
the other hand, median statistics, as opposed to arithmetic
means and RMS, proved to be robust enough against outliers.

If one plots the photometric scatter $\sigma_{\langle m \rangle}$ and $\sigma_{\langle m' \rangle}$ 
as a function of magnitude (Fig.~\ref{rmsq}, upper left panel, red vs.~green points), 
it is clear that $\sigma_{\langle m'\rangle} < \sigma_{\langle m \rangle}$ especially on bright 
stars ($-13.5< \mathrm{F467M} < -10$). The average  decrease in RMS is 
up to 15\% at $\mathrm{F467M}\simeq -13$ (Fig.~\ref{rmsq}, lower left panel).  
GZP-correction is therefore effective when compared to raw photometry. Still,
upon visual inspection spatial- and magnitude-dependent systematics are still present 
on the bright side of our sample and require a more
sophisticated correction. 

\subsection{Local differential photometry}

In order to build a set of suitable
reference light curves to evaluate a LZP correction, one needs
to trim down the list of comparison stars by rejecting variables and 
badly-behaved sources. First we considered the distribution of 
GZP-corrected scatter $\sigma_{\langle m'\rangle}$ as a function 
of $\langle m'_{i,j} \rangle_i$, evaluating its median and scatter over
magnitude bins; then each star more than $4\sigma$ off the median 
$\sigma_{\langle m'\rangle}$ was discarded from the reference set. The 
``loose'' $4\sigma$ threshold is justified by the need of not rejecting stars 
which could share common systematics with respect to other target stars;
in that case their inclusion in the reference set would be the only way to correct
such systematics.

For each pair of target star $k$ and reference star $j$, we 
constructed a differential light curve by subtracting their
raw magnitudes $m_{i,k}$ and $m_{i,j}$ on each frame $i$ where 
both stars are detected. Then 
we considered the distribution of the absolute 
residuals around $\langle m_{i,k}-m_{i,j}\rangle_i$, and define
the scatter $\sigma_{jk}$ as the $68.27^\mathrm{th}$ 
percentile of such residuals.  
The quantity $\sigma_{jk}$ is an empirical 
estimate of how much the reference star $j$ is a ``good'' reference for the 
target $k$. We can then assume the quantities $w_{jk}=1/\sigma^2_{jk}$ as 
initial weights to compute a more accurate ZP correction for a given target $k$. 

Since we want the correction to
be \emph{local}, we also multiply the weights by a factor $D_{jk}$ which is dependent
on the relative on-sky position $\varrho^2_{jk} = (x_j-x_k)^2 + (y_j-y_k)^2 $
between the reference star ($x_j$, $y_j$) and the target star ($x_k$, $y_k$).
To avoid using a reference star too close to the target, 
which could therefore be blended or contaminated,
$D_{jk}$ is forced to zero within $r_0$. Outside $r_0$,
we chose to parametrise $D_{jk}$ as a unitary factor up to a inner radius $r_\mathrm{in}$,
and then as a smooth function which decreases exponentially from one to zero with a 
$r_\mathrm{out}-r_\mathrm{in}$ scale radius (i.e., the factor $D_{jk}$ is 
$1/\mathrm{e}$ at $r_\mathrm{out}$): 
\begin{equation}
D_{jk} = 
\left\{ 
\begin{array}{cl}
0 & \textrm{if } \; \varrho_{jk} < r_0 \\
1 & \textrm{if } \; r_0 \leq \varrho_{jk} \leq r_\mathrm{in} \\
\exp \left [-\left ( \frac{\varrho_{jk}-r_\mathrm{in}}{r_\mathrm{out}-r_\mathrm{in}}\right )^2 \right ]  & \textrm{if } \; \varrho_{jk} > r_\mathrm{in}\textrm{ .} \\
\end{array}
\right .
\end{equation}
A similar weight factor $M_{jk}$ is imposed on the magnitude difference 
$\phi_{jk}=|m_j-m_k|$, as we expect that systematics due to non-linearity and background 
estimation are magnitude-dependent. 
The flux boundaries are
$f_\mathrm{in}$ and $f_\mathrm{out}$, respectively:
\begin{equation}
M_{jk} = 
\left\{ 
\begin{array}{cl}
1 & \textrm{ if } \; \phi_{jk} \leq f_\mathrm{in}  \\
\exp \left [-\left ( \frac{\phi_{jk}-f_\mathrm{in}}{f_\mathrm{out}-f_\mathrm{in}}\right )^2 \right ] & \textrm{ if } \; \phi_{jk} > f_\mathrm{in} \textrm{ .}  \\
\end{array}
\right .
\end{equation}

Summarising, the final weights are given by multiplying the above factors:
\begin{equation}
W_{jk} = \left ( 1/\sigma^2_{jk} \right ) \cdot D_{jk} \cdot M_{jk} \textrm{ .}
\end{equation}
The LZP correction $\tau'_i$ is evaluated as for the GZP correction 
(Eq.~\ref{gzp}), but this time using 
the weighted mean of magnitudes of our set of reference stars instead of an 
unweighted median, where the weights are assigned as $W_{jk}$. A
3$\sigma$ clip is applied on each image $i$ to improve robustness.

Our approach gives larger weights to reference stars which 1) produce
a smaller scatter on the target light curve; 2) 
are geometrically closer to the target as projected on the sky; 3) have a 
magnitude similar to that of the target. Of course this approach could
be easily generalised by introducing weights based on other external
parameters, such as colour or background level, for instance.

The five input parameters
$r_0$, $r_\mathrm{in}$, $r_\mathrm{out}$, $f_\mathrm{in}$, $f_\mathrm{out}$ have to 
be chosen empirically. After some iterations, we set $r_0=20$ pix,
$r_\mathrm{in}=200$ pix, $r_\mathrm{out}=300$ pix, $f_\mathrm{in}=1.0$ mag, 
$f_\mathrm{out}=1.75$ mag. The improvement of the LZP correction over the GZP one
is shown in the left panel of Fig.~\ref{rmsq}. In the bright, non-saturated
end of our sample ($\mathrm{F467M}\simeq -13$) the RMS is lowered by 10-25\%
on average when compared to GZP-corrected and raw photometry, respectively.
On the faintest targets, GZP performs slightly better than LZP because 
photon noise dominates and GZP is not forced to discard very bright stars 
as LZP does. For each target, we chose to apply an \emph{optimal} correction 
which outputs the light curve having the lowest scatter
among the raw one and the GZP- or LZP-corrected ones. From here on all
procedures are carried out on such optimal light curves.

\subsection{Variable-searching procedures}

A battery of software tools to detect both periodic and non-periodic
photometric variability was applied to all $9\,410$ corrected
light curves. These tools included period-searching algorithms such as
the classical Lomb \& Scargle periodogram (LS; \citealt{lomb1976}; \citealt{scargle1982}) and its 
generalised version GLS \citep{zechmeister2009}; the Analysis 
of Variance periodogram (AoV; \citealt{schwarzenberg1989}); the Box-fitting 
Least-Squares periodogram  (BLS; \citealt{kovacs2002}). The latter is the
most sensitive to eclipse-like event, such as those expected from detached 
eclipsing binaries and planetary transits. A second class of
diagnostics was exploited to search for more general types of
variability: the alarm variability statistic as described by \citet{tamuz2005}, 
the overall scatter (based on robust median statistics as defined at the
beginning of Sect.~\ref{ssgzp}) as a function of magnitude, and the RMS 
after each of the 12 astrometric epochs has been averaged on a single bin, 
to catch the effects of long-term variability.

\begin{figure}
\includegraphics[width=\columnwidth]{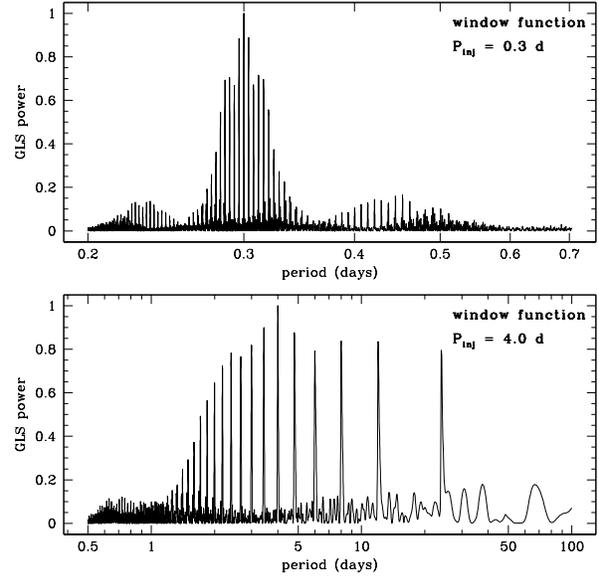}
\caption{Noise-free window function computed over the temporal baseline of our data set;
the injected periods are $P_\mathrm{inj}=0.3$ days (\emph{upper panel}) 
and $P_\mathrm{inj}=4.0$ days (\emph{lower panel}).}
\label{window}
\end{figure}

We already mentioned that our data set is not optimised to search for
periodic variability. One of the most limiting factors is the 
non-regular cadence, as each 
astrometric epoch is separated by 24 days. These temporal 
gaps introduce many spurious frequencies in the periodograms,
making the recovery of the true (astrophysical) period problematic,
especially for signals at $P<24$ d where most of the variables are
expected to be. We illustrate this by injecting noise-free sinusoids
over the 589-images time 
baseline of our WFC3 data set, and then
recovering the signal through a GLS periodogram on the synthetic light
curve (Fig.~\ref{window}). In both cases at $P_\mathrm{inj}=0.3$ and
4.0 days, we get a ``comb'' of periodogram peaks instead of a sharp
spike, as it would be expected in an optimal sampling regime. When
random noise and systematic errors are accounted for, period
recovering gets harder, and the 24-day alias induced by sampling
becomes the most significant period. For this reason we split our
search over two period ranges: 0.1-20 and 20-200 days, analysing each
set independently.

To visually supervise all the individual outputs of the analysis described above on more
than 9\,000 targets would be much too time-consuming and prone to biases. Instead, we 
selected a shortlist of candidate variables by running the very same 
analysis on a set of synthetic light curves, sampled at the same epochs as
the real data but after having randomly shuffled the magnitude values. 
In this way, noise and sampling cadence are preserved, while phase coherence is
broken: the resulting ``synthetic'' analysis represents the expected output when 
an intrinsic signal is not present. We focused on the distribution of 
diagnostics such as:
\begin{enumerate}
\item the periodogram \emph{power} for the LS, GLS and AoV 
algorithm (as defined in the original papers, i.e., $P_X$
by \citealt{scargle1982}, $p(\omega)$ by \citealt{zechmeister2009}, and $\Theta$
by \citealt{schwarzenberg1989}, respectively);
\item the \emph{SN} and \emph{SDE} statistics \citep{kovacs2002} and
the signal-to-pink noise ratio (as defined by \citealt{pont2006} and
implemented by \citealt{hartman2008}) for the BLS periodogram;
\item the alarm index $\mathcal{A}$ \citep{tamuz2005} and the visit-binned and 
unbinned RMS as a function of magnitude, defined as above.
\end{enumerate}
For each of the above diagnostics, the output distribution from the real
data was compared with the results from the synthetic light curves, selecting
all targets which fall at least 3-$\sigma$ outside the latter distribution. 
This gave us a list of 401 candidates which were individually inspected.
Most of them turned out to be spurious, due to the target being blended,
contaminated, or part of an extended source. 
Very often false 
positives came from sources with bad photometry on a 
single visit due to the star falling on a bad pixel or too close to the 
detector edges. The latter case was the most frequent cause of false
positives detected at periods around the 12- or 24-day alias.

\section{Results}

After the vetting process, only 38 variables survived. They are listed 
in Table \ref{variables}. All of them appear to be
isolated, point-like sources, following a visual 
inspection of the images. This is also confirmed by their 
\texttt{qfit} diagnostic, which is perfectly consistent with 
the median \texttt{qfit} of well-measured stars of similar magnitude
(Fig.~\ref{rmsq}, right plot; variables are marked with red stars). Their 
detected periods are clearly distinct from the typical periods due to 
aliasing or instrumental effects, such as the 96-min orbital period of \emph{HST}
or the 24-day separation between consecutive visits. For the reasons above, we
can identify all those 38 stars as genuine astrophysical variables.
In Table \ref{variables} we also report the Johnson $V$-band magnitude of each
variable (obtained by cross-matching our catalogue with that by
\citealt{sarajedini2007}) and we identify which stars are matched within a 
2-$\sigma$ error ellipse with an X-ray source from the \citet{bassa2004}
catalogue.
We stress out that the reported  $V$ magnitudes represent the
average of the values obtained by Sarajedini et al., not an
intensity-weighted average throughout the phase of our
light curves.

\begin{figure*}
\includegraphics[width=2\columnwidth]{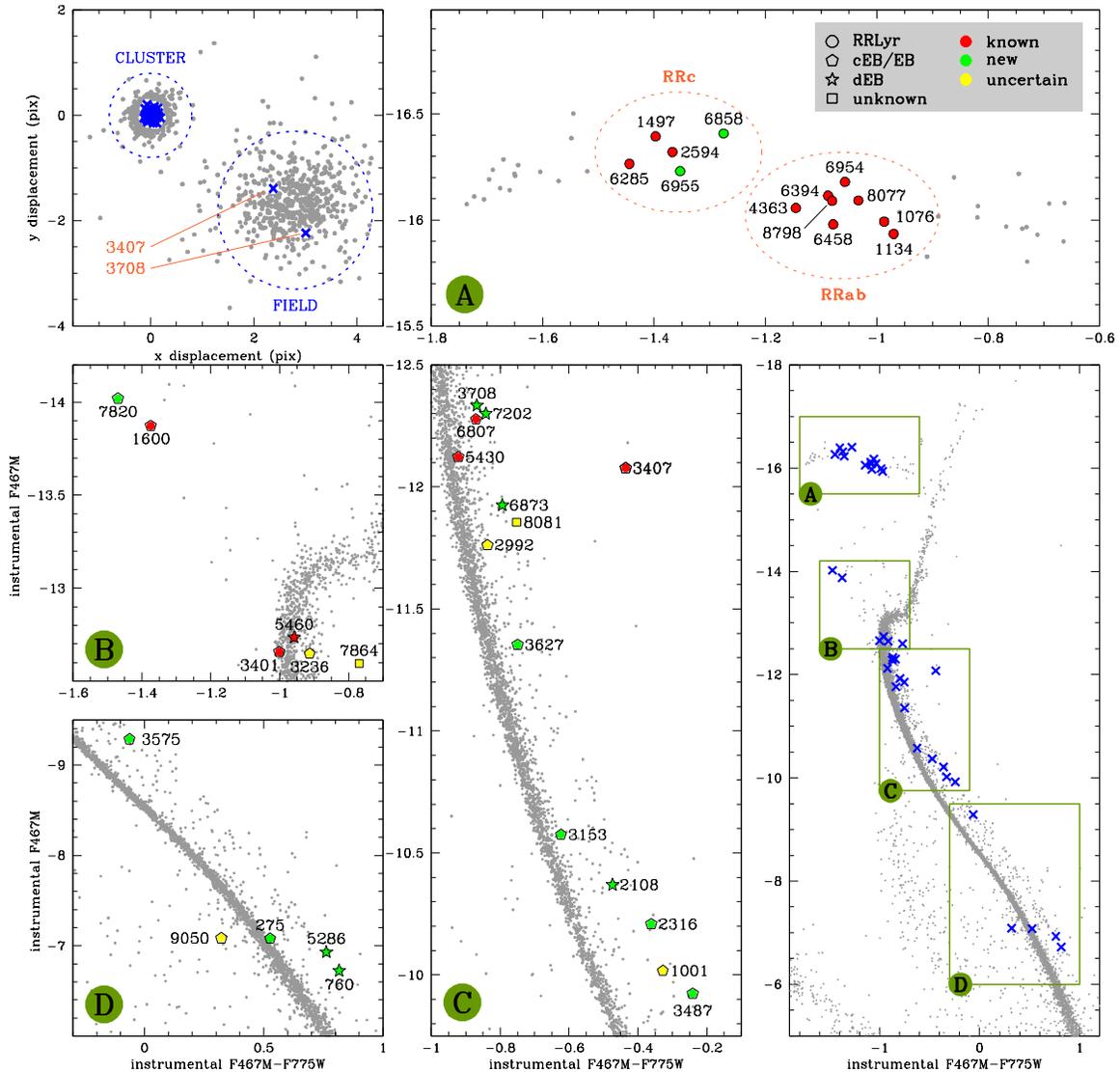}
\caption{Proper motion vector--point and colour--magnitude diagrams (VPD, CMDs) 
of M\,4 where the detected variables have been highlighted. \emph{Upper left}: $x$ and $y$ proper motion
displacements in physical pixels, for each analysed star (grey dots) and detected variables (blue crosses). 
Among the latter, only ID\#3407 and 3708 appear to not be cluster members. \emph{Lower
right}: instrumental F467M, F775W CMD of M\,4 from our data set. Variables are marked
with blue crosses. Regions within green boxes ($A$, $B$, $C$, $D$) are zoomed and displayed in the
corresponding remaining panels, and labelled according to the legend. 
}
\label{cmd}
\end{figure*}

The membership of our variables with respect to M\,4 can be assessed with a very high
level of confidence by inspecting the proper motion vector-point diagram (VPD), where
stars belonging to the cluster and those in the general field appear extremely well 
separated (Fig.~\ref{cmd}, upper left plot). 
To our purposes the VPD does not need to be calibrated in physical 
units ($''/\mathrm{yr}$ proper motion); instead, we simply plot the displacement
in pixels measured over the baseline between the WFC3 observations
and the first astrometric ACS epoch ($\sim$6 years).
We found that nearly all variables are cluster members 
with the only exceptions being ID\#\ 3407 and 3708. As expected, most variables 
found at magnitudes fainter than the cluster turnoff turned out to be eclipsing binaries, both of
contact (cEB) and detached (dEB) subtypes. On most cases this classification 
is also supported by their position on the CMD,
which is shifted up to $\sim$0.75 mag upwards with respect to single, unblended main sequence stars
(Fig.~\ref{cmd}). A discussion of individual cases follows.

\subsection{Notes on individual objects}

\noindent\emph{Known RR Lyrae: ID\#\ 1076,\ 1134,\ 1497,\ 2594,\ 4363,\ 6285,\ 6394,\ 6458,\ 6954,\ 8077,\ 8798}. 
These are RR Lyr variables known since a long time (\citealt{greenstein1939}; re-identified by
\citealt{shokin1996}), but whose periods are here determined with much more precision given
the 1-yr temporal baseline. The discrimination between RRab (ID\# 1076, 1134, 4363, 6394, 6458, 6954, 8077, 8798) 
and RRc (1497, 2594, 6285) subtypes is obvious. Our classification is
confirmed by their position in the CMD (Fig.~\ref{cmd}, right panels), 
with RRab and RRc member being clearly separated by the RR Lyrae gap.
The amplitudes of ID\# 6285 and 8077 change significantly through the series,
due possibly to the Blazhko effect \citep{kovacs2009}.

\medskip

\noindent\emph{New RR Lyrae: ID\#\ 6858,\ 6955}. These are two very close ($1.6''$) 
RR Lyr variables having similar magnitude ($V=13.24$ vs.~$13.39$).
Such an unusual pair is mostly blended on ground-based images, so it is not surprising that it was
classified by \citet{greenstein1939} as a single RR Lyr with a problematic light 
curve (C40) and a poorly-constrained period. Other studies recognised it as a visual binary 
but failed at discovering the true nature of both sources \citep{desitter1947}; 
therefore C40 has been neglected 
in many follow-up works on RR Lyrae. We identified both stars 
as RRc subtypes with periods $P\simeq0.39$ and $0.29$ d, respectively. 

\medskip

\noindent\emph{Blue stragglers: ID\#\ 1600 and 7820} are without any doubt cluster
members based on their proper motions, and are located in the ``blue straggler'' region of 
the CMD. ID\#\ 1600 was already known as a short-period, near equal-mass
contact eclipsing binary \citep{kaluzny1997}. 
ID\#\ 7820 also is a contact binary, reported here
for the first time. Its periodic modulation at $P\simeq 0.66$ d
is detected at high significance. Its primary and 
secondary minima, showing very unequal depths, suggest a much lower 
mass ratio than ID\#\ 1600.

\medskip

\noindent\emph{Known contact EBs}. ID\#\ 3401,\ 3407,\ 5430,\ 6807 are
already listed in the \citet{clement2001} catalogue. They 
are located close to the turnoff region,
with very well defined primary and secondary minima and periods 
spanning 0.26-0.30 d. Among these, ID\# 3407 is the only 
field star, clearly separated from the cluster in the PM diagram;
also it is an X-ray source catalogued as CX13 by \citet{bassa2004}.

\medskip

\noindent\emph{Detached EBs: ID\#\ 760,\ 2108,\ 3708,\ 5286,\ 5460,\ 6873,\ 7202}
are detached eclipsing binaries, all identified as cluster members by PMs
with the only exception of ID\# 3708. Among them only ID\# 5460 was previously 
published (as K66, by \citealt{kaluzny2013b}). All of them lie in the upper
part of the binary main sequence, i.e., are systems with high mass
ratios ($q\approx 1$). 
For that reason their phased light curves are expected to show two
eclipses of similar depth between phases 0 and 1. But the
period-search algorithms does not know this and therefore might find
periods of half the true duration with light curves showing only one
eclipse between phases 0 and 1. In some cases, such as for ID\# 2108, we
have no way of knowing the true period with the current data because
of holes in the phased light curves where an eclipse might be
happening. In other cases, such as for ID\# 5286, the phase coverage seems
sufficient to rule out the presence of a second eclipse at the period
found, which suggests that the true period is twice as long. Based on
such considerations, we give the most likely true periods in Table 1. 
%%%
ID\# 2108 also is an X-ray source 
(CX 28; \citealt{bassa2004}), but its light curve shows no sign of 
stellar activity or interaction.
We set the period of ID\# 3708  as a lower limit ($P>6.47$ d) since only one eclipse is
detectable in our series.

\medskip

\noindent\emph{New contact and generic EBs: ID\#\ 2316,\ 3153,\ 3487,\ 3575}.  
The light curves of 
ID\#\ 2316 and 3153 show an anomalous amount of scatter despite being isolated 
and well-measured (low \texttt{qfit}); this 
could be ascribed to starspot-induced variability by one or both components
of the eclipsing binary. 
ID\#\ 3487 is a much fainter EB in the lower MS, observed at low S/N but with 
recognisable minima and maxima.

\medskip

\noindent\emph{Probable EBs: ID\#\ 1001,\ 2992,\ 3236,\ 3627}. 
ID\#\ 1001 shows a sharp $\sim$0.1-mag eclipse overimposed on a pseudo-sinusoidal
modulation. If the modulation is interpreted as caused by 
a ``hot spot'' or persistent active region 
then the eclipse appears to be out of phase by about 0.25 
(assuming circular orbits, and spin-orbit alignment).
No secondary minimum is detected. Light curves of 
ID\#\ 2992, 3236, 3627 share similar amplitudes (0.05-0.1 mag) and 
an excess of photometric scatter, 
but their position in the CMD makes a cEB/dEB classification plausible. 
ID\# 3236 (=CX3) and 3627 (=CX20) are also matched to X-ray
sources.

\medskip

\noindent\emph{Unclassified/uncertain: ID\#\ 275,\ 7864,\ 8081,\ 9050}.
The light curve of ID\# 275 is clearly periodic with a ``double-wave''
shape and minima of similar depths. Anyway, its period $P\simeq 0.18$ d is smaller 
than the usual $\sim$0.22 d cut-off found for MS+MS eclipsing
binaries \citep{norton2011} and its position on the CMD is 
very close to the MS ridge, disproving the presence of companions with 
high $q$. On the other hand, ID\# 275 could be a new member of the
recently discovered class of ultra-short period M dwarf binaries, 
first introduced by \citet{nefs2012}. 
ID\# 7864 is a known variable and X-ray
source (K52; CX8), and is much brighter than the upper binary sequence envelope, 
revealing itself as an interacting and/or higher-multiplicity system. This is confirmed by
its light curve, showing a strong $P\simeq 0.77$ d periodicity but also a heavily 
disturbed signal, as already noticed by \citet{kaluzny2013b}. 
A similar light curve is detected at $P\simeq 0.97$ d on the
previously uncatalogued ID\# 8081, also a X-ray source (CX11). ID\#
9050 is even more peculiar: just as ID \# 275, the signal is a clean
``double-wave'' at $P\simeq 0.18$ d, again an unusually short period for a cEB, but
furthermore its position on the CMD is slightly \emph{bluer} than the
MS, suggesting a possible binarity with a bright WD.  
Only a targeted follow-up could reveal more about the nature of
this target.

\begin{figure*}
\includegraphics[width=2\columnwidth]{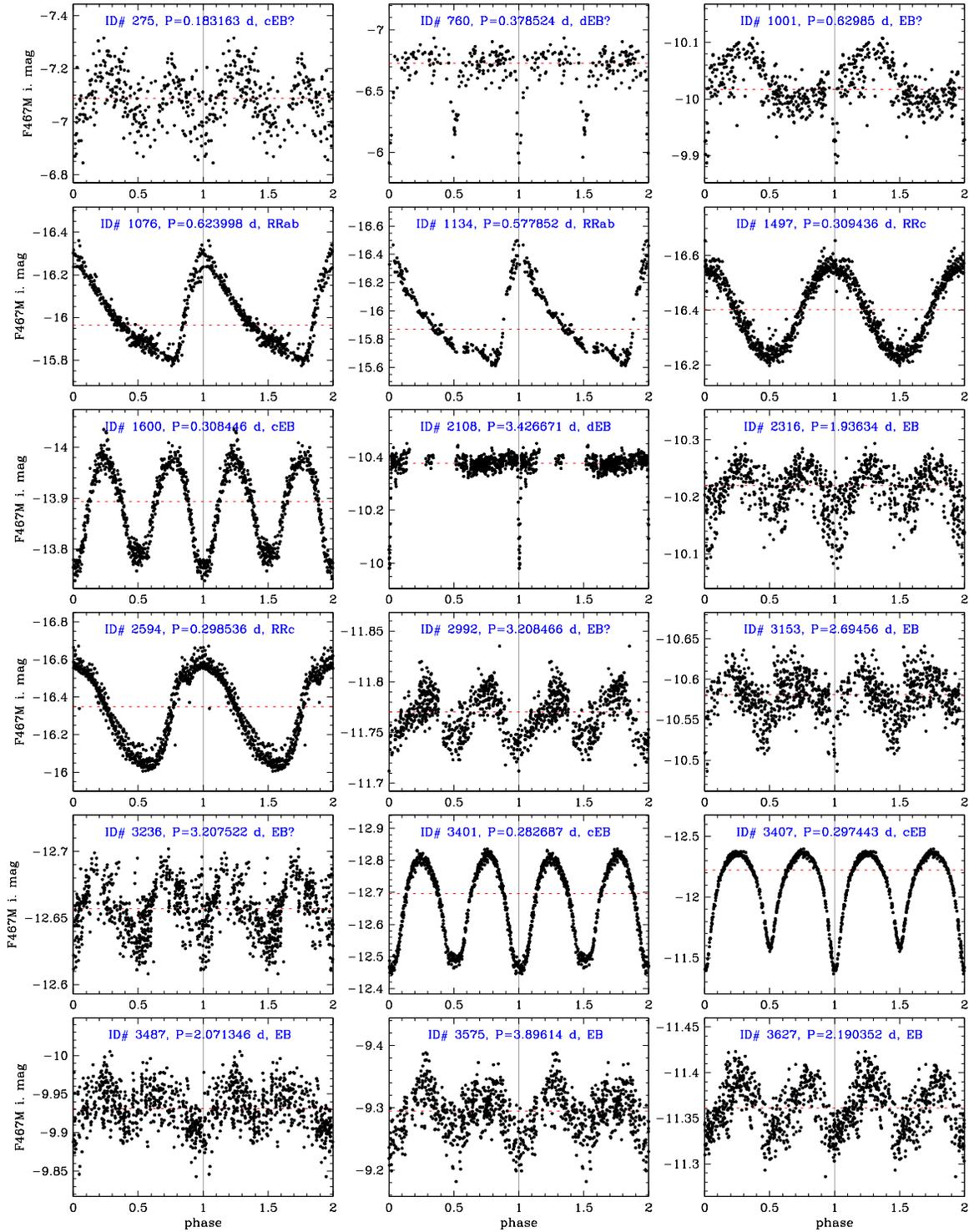}
\caption{Light curves of variables detected on our data set, folded on the 
most significant period found by periodograms (\emph{figure 1 of 2}). 
Data points are repeated twice over phase 
for clarity. Target are sorted by ID\# code. See Table \ref{variables} for coordinates and cross-referencing.
}
\label{lcs1}
\end{figure*}

\begin{figure*}
\includegraphics[width=2\columnwidth]{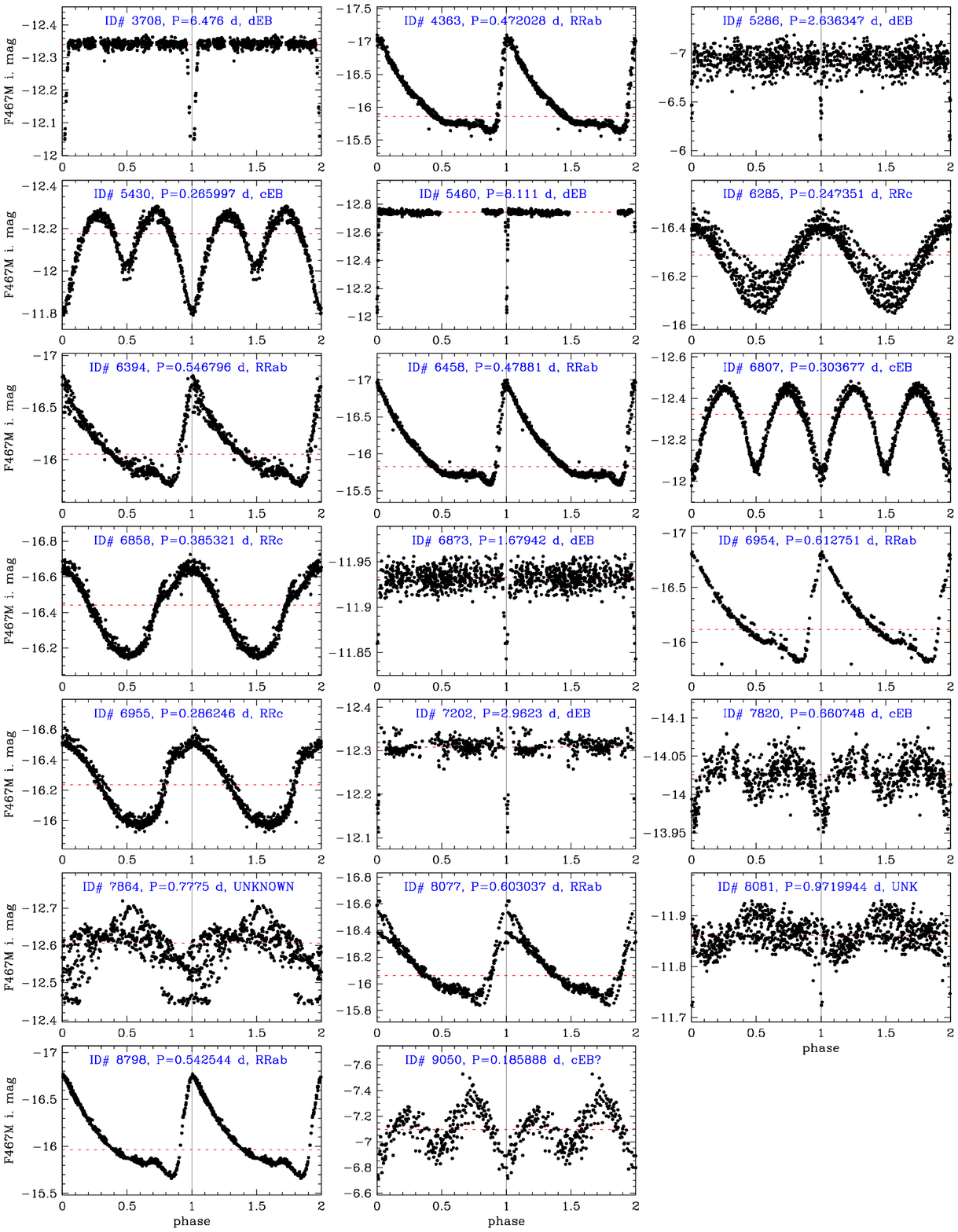}
\caption{Light curves of variables detected on our data set, folded on the 
most significant period found by periodograms (\emph{figure 2 of 2}). Data points are repeated twice over phase 
for clarity. Target are sorted by ID\# code. See Table \ref{variables} for coordinates and cross-referencing.}
\label{lcs2}
\end{figure*}

\begin{table*}
\caption{Variables found and their characteristics.}
\medskip
\label{variables}
\begin{tabular}{rrrrrrlllll}
\hline
  \multicolumn{1}{c}{$N$} &
  \multicolumn{1}{c}{ID\#} &
  \multicolumn{1}{c}{$\alpha$~(J2000)} &
  \multicolumn{1}{c}{$\delta$~(J2000)} &
  \multicolumn{1}{c}{F467M} &
  \multicolumn{1}{c}{$V$} &
  \multicolumn{1}{c}{M} &
  \multicolumn{1}{c}{Type} &
  \multicolumn{1}{c}{$P$~(days)} &
  \multicolumn{1}{c}{X-ref} &
  \multicolumn{1}{l}{Notes} \\
\hline\hline
  1  & 275  & 245.90461 & $-$26.55051 & $-$7.08  & 21.77 & Y & cEB  & 0.183163 & -- & MS\\
  2  & 760  & 245.88814 & $-$26.55350 & $-$6.65  & 22.03 & Y & dEB  & 0.378524 & -- & BSEQ\\
  3  & 1001 & 245.92283 & $-$26.53479 & $-$10.02 & 19.30 & Y & EB?  & 0.62985  & -- & BSEQ\\
  4  & 1076 & 245.89437 & $-$26.54783 & $-$16.00 & 13.56 & Y & RRab & 0.623998 & C39 & HB\\
  5  & 1134 & 245.88684 & $-$26.55099 & $-$15.94 & 13.91 & Y & RRab & 0.577852 & C38 & HB\\
  6  & 1497 & 245.89738 & $-$26.54333 & $-$16.40 & 13.30 & Y & RRc  & 0.309436 & C20 & HB\\
  7  & 1600 & 245.91036 & $-$26.53634 & $-$13.88 & 15.74 & Y & cEB  & 0.308446 & C72=K53 & BSS\\
% 8  & 2108 & 245.89580 & $-$26.54005 & $-$10.36 & 18.96 & Y & dEB  & 3.426671 & -- & BSEQ; CX28\\
  8  & 2108 & 245.89580 & $-$26.54005 & $-$10.36 & 18.96 & Y & dEB  & $6.853342^{\ddagger?}$ & -- & BSEQ; CX28\\
  9  & 2316 & 245.88963 & $-$26.54161 & $-$10.21 & 19.17 & Y & EB   & 1.93634  & -- & BSEQ; CX25\\
  10 & 2594 & 245.90545 & $-$26.53228 & $-$16.32 & 13.05 & Y & RRc  & 0.298536 & C23 & HB\\ \hline
% 11 & 2992 & 245.89616 & $-$26.53445 & $-$11.76 & 17.64 & Y & EB?  & 1.604233 & -- & MS/BSEQ\\
  11 & 2992 & 245.89616 & $-$26.53445 & $-$11.76 & 17.64 & Y & EB?  & 3.208466 & -- & MS/BSEQ\\
  12 & 3153 & 245.89446 & $-$26.53451 & $-$10.57 & 18.91 & Y & EB   & 2.69456  & -- & MS/BSEQ; CX21\\ %CX?
% 13 & 3236 & 245.90870 & $-$26.52718 & $-$12.65 & 16.89 & Y & EB?  & 1.603761 & -- & BSEQ/TO; CX3?\\
  13 & 3236 & 245.90870 & $-$26.52718 & $-$12.65 & 16.89 & Y & EB?  & 3.207522 & -- & BSEQ/TO; CX3\\ %CX?
  14 & 3401 & 245.90329 & $-$26.52891 & $-$12.66 & 16.79 & Y & cEB  & 0.282687 & C67=K48 & TO; CX15\\ %CX?
  15 & 3407 & 245.89302 & $-$26.53385 & $-$12.08 & 16.98 & N & cEB  & 0.297443 & C68=K49 & NM; CX13\\
  16 & 3487 & 245.90270 & $-$26.52871 & $-$9.93  & 19.30 & Y & EB   & 2.071346 & -- & BSEQ\\
  17 & 3575 & 245.91544 & $-$26.52222 & $-$9.29  & 19.85 & Y & EB   & 3.89614  & -- & BSEQ\\
% 18 & 3627 & 245.90375 & $-$26.52755 & $-$11.36 & 18.09 & Y & EB   & 1.095176 & -- & MS/BSEQ; CX20\\
  18 & 3627 & 245.90375 & $-$26.52755 & $-$11.36 & 18.09 & Y & EB   & 2.190352 & -- & MS/BSEQ; CX20\\
  19 & 3708 & 245.90634 & $-$26.52592 & $-$12.33 & 17.17 & N & dEB  & $>$6.476 & -- & NM; one eclipse\\
  20 & 4363 & 245.89960 & $-$26.52604 & $-$15.97 & 13.70 & Y & RRab & 0.472028 & C21 & HB\\ \hline
% 21 & 5286 & 245.89808 & $-$26.52231 & $-$6.92  & 21.87 & Y & dEB  & 2.636347 & -- & BSEQ\\
  21 & 5286 & 245.89808 & $-$26.52231 & $-$6.92  & 21.87 & Y & dEB  & $5.272694^\ddagger$ & -- & BSEQ\\
  22 & 5430 & 245.88048 & $-$26.53014 & $-$12.13 & 17.53 & Y & cEB  & 0.265997 & C69=K50 & MS\\
  23 & 5460 & 245.88432 & $-$26.52812 & $-$12.73 & 16.80 & Y & dEB  & 8.111    & K66 & TO\\
  24 & 6285 & 245.88155 & $-$26.52519 & $-$16.27 & 13.60 & Y & RRc  & 0.247351 & C37 & HB\\
  25 & 6394 & 245.90840 & $-$26.51161 & $-$16.12 & 13.36 & Y & RRab & 0.546796 & C24 & HB\\
  26 & 6458 & 245.89448 & $-$26.51795 & $-$15.92 & 13.72 & Y & RRab & 0.47881  & C18 & HB\\
  27 & 6807 & 245.88831 & $-$26.51917 & $-$12.28 & 17.08 & Y & cEB  & 0.303677 & C70=K51 & BSEQ\\
  28 & 6858 & 245.90195 & $-$26.51230 & $-$16.41 & 13.24 & Y & RRc  & 0.385321 & C40$\dagger$ & HB\\
% 29 & 6873 & 245.89321 & $-$26.51636 & $-$11.93 & 17.56 & Y & dEB  & 1.679423 & -- & BSEQ\\
  29 & 6873 & 245.89321 & $-$26.51636 & $-$11.93 & 17.56 & Y & dEB  & $3.358846^\ddagger$ & -- & BSEQ\\
 30 & 6954 & 245.91415 & $-$26.50592 & $-$16.18 & 13.07 & Y & RRab & 0.612751 & C25 & HB\\ \hline
  31 & 6955 & 245.90169 & $-$26.51190 & $-$16.23 & 13.39 & Y & RRc  & 0.286246 & C40$\dagger$ & HB\\
% 32 & 7202 & 245.86887 & $-$26.52611 & $-$12.30 & 17.20 & Y & dEB  & 2.962981 & -- & BSEQ\\
  32 & 7202 & 245.86887 & $-$26.52611 & $-$12.30 & 17.20 & Y & dEB  & $5.925962^{\ddagger?}$ & -- & BSEQ\\
  33 & 7820 & 245.90261 & $-$26.50605 & $-$14.02 & 15.76 & Y & cEB  & 0.660748 & -- & BSS\\
  34 & 7864 & 245.88111 & $-$26.51606 & $-$12.58 & 16.92 & Y & UNK  & ---      & C71=K52 & above BSEQ; CX8\\
  35 & 8077 & 245.90387 & $-$26.50362 & $-$16.09 & 13.17 & Y & RRab & 0.603037 & C22 & HB\\
  36 & 8081 & 245.88501 & $-$26.51263 & $-$11.86 & 17.65 & Y & UNK  & 0.971994 & -- & BSEQ; CX11\\
  37 & 8798 & 245.88532 & $-$26.50639 & $-$16.09 & 13.55 & Y & RRab & 0.542544 & C16 & HB\\
  38 & 9050 & 245.89351 & $-$26.49984 & $-$7.09  & 21.95 & Y & cEB?  & 0.185888 & -- & under the MS\\
\hline\end{tabular}

\justify
\bigskip
\textbf{Notes.} The columns give: a progressive number $N$,
the ID code of the source according to our catalogue, the equatorial
coordinates $\alpha$ and $\delta$ at epoch 2000.0, the instrumental
magnitude in F467M and the standard Johnson $V$ magnitude from \citet{sarajedini2007}, 
a membership flag derived from the VPD,
the variability class assigned by our study, the most probable 
photometric period $P$ (where detectable), the cross-references
to the variable catalogues by \citet{kaluzny2013b} (letter K)
and \citet{clement2001} (letter C), and some notes. Notes are
codified as follows: MS = main sequence star, BSEQ = binary sequence star 
(between fiducial MS and MS+0.75 mag), HB = horizontal branch, BSS =
blue straggler star, CX = X-ray source from the catalogue by \citet{bassa2004}, 
TO = turn-off star, NM = not a cluster member.
A few periods of dEBs, marked with a $\ddagger$, have been doubled with respect
to the best-fit periodogram solution, following astrophysical arguments (see Sec.~3.1
for details); two cases are ambiguous, and marked with an additional question mark.
The two variables marked by $\dagger$ in the X-ref field were once classified as a single,
atypical RR Lyr (\citealt{clement2001} and references therein).
\end{table*}

\begin{table*}
\caption{Light curves of the variables found.}
\medskip
\label{lcstable}

\begin{tabular}{rrrrrrr}
\hline
  \multicolumn{1}{c}{ID\#} &
  \multicolumn{1}{c}{BJD(TDB)} &
  \multicolumn{1}{c}{F467M} &
  \multicolumn{1}{c}{$\Delta$F467M} &
  \multicolumn{1}{c}{$x$} &
  \multicolumn{1}{c}{$y$} &
  \multicolumn{1}{c}{\texttt{qfit}} \\
\hline\hline
1001 &  2456209.701194 & -10.0180 &   0.0168 &   2653.966 &   2571.443 &  0.037 \\
1001 &  2456209.740334 & -10.0160 &   0.0168 &   2653.973 &   2571.472 &  0.046 \\
1001 &  2456209.746422 & -10.0060 &   0.0168 &   2653.999 &   2571.477 &  0.032 \\
1001 &  2456209.752509 & -10.0130 &   0.0168 &   2653.973 &   2571.466 &  0.039 \\
1001 &  2456209.764684 &  -9.9920 &   0.0168 &   2653.994 &   2571.487 &  0.031 \\
1001 &  2456209.806845 & -10.0040 &   0.0168 &   2654.001 &   2571.439 &  0.033 \\
1001 &  2456209.812933 & -10.0110 &   0.0168 &   2653.990 &   2571.455 &  0.037 \\
1001 &  2456209.819020 & -10.0080 &   0.0168 &   2653.991 &   2571.449 &  0.039 \\
1001 &  2456209.825108 & -10.0200 &   0.0168 &   2653.965 &   2571.461 &  0.037 \\
1001 &  2456209.831195 & -10.0035 &   0.0168 &   2653.963 &   2571.475 &  0.048 \\ 
\hline\end{tabular}

\justify
\bigskip
\textbf{Notes.} Table \ref{lcstable}
is published in its entirety as a machine-readable table in the online
version of this article and the Centre de Donees Strasbourg (CDS). A
portion is shown here for guidance regarding its form and content.
The columns give: the ID code of the source according to our catalogue (see Table \ref{variables}), 
the mid-exposure time in the BJD-TDB time standard \citep{eastman2010}, 
the instrumental magnitude in F467M and the associated photometric error,
the $x$ and $y$ centroid position on the chip, and the \texttt{qfit}
quality parameter. 
\end{table*}

\section{Discussion and Conclusions}

In the previous sections we described how we performed a 
search for photometric variability among a sample of 9410 
stars in a field imaged by \hst{} on the core of the globular
cluster M\,4. Such a search yielded 38 variable stars;
all but two are cluster members, and 20 are reported here
for the first time. Quite surprisingly, two newly (re-)classified
sources are a pair of bright but blended RR Lyrae, whose true nature
has been unveiled by the superior angular resolution of
space-based imaging. A few candidates cannot be assigned to
standard variability classes, being aperiodic, multiperiodic, 
or lying in unusual regions of the CMD. 

We did not detect any signal which could be interpreted as a 
transit by planet-sized objects orbiting around solar-type stars 
(i.e., box-shaped 
eclipses having photometric depth smaller than 0.03 mag). 
Many intrinsic factors in our data set are strongly limiting
a transit search, above all sparse temporal sampling 
(which decreases phase coverage, and worsens the effects of long-term
stellar variability) and large-scale 
dithering (which boosts position-dependent systematic errors).
% It's too bad we don't live on Jupiter.
As only planets among the relatively rare ``hot Jupiter'' class are 
within reach of such a search, the statistical significance
of our null detection is probably very low. We will 
investigate this further in a forthcoming paper of the ``M\,4 core project''
series focused on transit search, which will also analyse 
time-series photometry from the parallel ACS fields.

The overall completeness level of our search is difficult to
quantify without posing very special assumptions about the 
shape and period of the photometric signal to be recovered. 
However, we note that we firmly detected variables having 
photometric amplitudes on the order of a few hundredths of 
magnitude even in the fainter half of our sample (such 
as ID\# 1001, 3487, 3575, 5286). Even though periodograms
are aliased at some characteristic frequencies 
by the particular sampling cadence of our data (Fig.~\ref{window}), 
phase coverage is complete up to periods of about six days. Merging 
the picture, this means that eclipsing contact binaries should be
detectable with a completeness factor close to one, with only
rare exceptions expected from very grazing
systems, or from binaries with mass ratios $q$ much smaller
than one.

About half of the reported variables (21, of which 19 
are high-confidence cluster members) are certain or probable eclipsing 
binaries. As we detected 
19 EBs among 5\,488 MS stars with $-12.75<\textrm{F467M}<-7.0$, or --restricting
the magnitude range to brighter targets-- 16 EBs among 4\,080 MS stars with 
$-12.75<\mathrm{F467M}<-9.0$, we estimate an \emph{observed} EBs fraction of about 0.3-0.4\%. 
This value is approximately consistent with the measured photometric
binary fraction in the core of $14.8\%\pm1.4\%$
(Milone et al.\ 2012), when allowance is made for the low
fraction of binaries which are expected to exhibit eclipses.  Using
population synthesis, \citet{soderhjelm2005}
show that the
fraction of all F and G stars exhibiting eclipses with a depth of a
few tenths of a magnitude and a period of a day is about 3\%.  The
population they studied had a binary fraction of 80\%, however, and so
the corresponding result for the core of M\,4 would be expected to be
about 0.5\%, i.e., close to our results.  A more precise
discussion depends on several factors, such as the details of the
period distribution of the observed binaries, dynamical erosion of the
binary population, and the initial binary distributions assumed in the
population synthesis.  Indeed the observations reported in this paper
are likely to be a useful constraint on our future dynamical modelling
of M\,4.

%%% 
As a final remark, we note that
the dEBs that are cluster members can
potentially be used for obtaining precise cluster parameters, insights
into multiple-populations of the cluster and stellar evolution tests
in general (paper I, \citealt{brogaard2012}, \citealt{milone2014}). This requires however that
the dEBs are bright enough for spectroscopic measurements. Based on
our experience from dEBs in the open cluster NGC6791, four of the dEBs
we identify in M\,4 (ID\# 5460, 7202, 6873, and 2108) are within reach of
current spectroscopic facilities such as UVES at the Very Large
Telescope. Adding also the additional dEBs found and analysed by
\citet{kaluzny2013a} makes a sample of six dEBs in the turn-off and
upper MS of M\,4 with great potential for improved cluster
insights. The two low-mass dEB systems found on the lower main
sequence ($V \simeq 22$; ID\# 760 and 5286) are too faint for
current spectroscopic facilities, but interesting potential targets for
a future ELT. We note that M\,4 will be in one of the fields of the K2
mission \citep{howell2014}, and that observing these six dEBs with K2
would solve period ambiguities and provide full-coverage light curves
valuable for their analysis.
%%%

\section*{Acknowledgements}

L.\ R.\ B., G.\ P., V.\ N., S.\ O., A.\ P. and L.\ M.\ acknowledge PRIN-INAF 2012
funding under the project entitled: ``The M\,4 Core Project with Hubble
Space Telescope''.
J.A., A.B., L.U. and R.M.R.\ acknowledge support from STScI grants GO-12911.
K.B.\ acknowledges support from the Villum Foundation.  
A.P.M.\ acknowledges the financial support from the Australian
Research Council through Discovery Project grant DP120100475.
M.\ G.\ acknowledges a partial support by the National Science Centre 
through the grant DEC-2012/07/B/ST9/04412.
V.~N.~acknowledges partial support from INAF-OAPd through
the grant ``Analysis of HARPS-N data in the framework of GAPS project''
(\#19/2013) and ``Studio preparatorio per le osservazioni della 
missione ESA/CHEOPS'' (\#42/2013).
Some tasks of our data analysis have been carried out with
the VARTOOLS (Hartman et al.~2008) and \texttt{Astrometry.net} codes
(Lang et al.\ 2010). This research made use of the International 
Variable Star Index (VSX) database, operated at AAVSO, Cambridge, 
Massachusetts, USA.
%%%%%%%%%%%%%%%%%%%%%%%%%%%%%%%%%%%%%%%%%%%%%%%%%%%%%%

%\nocite{*}

\bibliographystyle{mn2e}
\bibliography{biblio}

\bsp

\label{lastpage}

\end{document}